\documentclass[twocolumn]{aastex631}
\usepackage{natbib, color,subfigure,lineno}
\graphicspath{{./}{figures/}}

\newcommand{\hbeta}{H{$\beta$}\,}
\newcommand{\halpha}{H{$\alpha$}\,}

\newcommand{\CIII}{C\,{\tt III]}\,}

\newcommand{\CaII}{Ca\,{\tt II}\,}
\newcommand{\OII}{[O\,{\tt II}]\,3727}

\def\CIV{C\,{\tt IV}\,}
\def\MgII{Mg\,{\tt II}\,}

\def\OIII{[O\,{\tt III}]\,5007\,}

\def\SiIV{Si\,{\tt IV}\,}

\shorttitle{Improved Redshifts for DESI Quasars}
\shortauthors{Wu \& Shen}
\graphicspath{{./}{figures/}}

\begin{document}

\title{Improved Redshifts for DESI EDR Quasars}

\author[0000-0003-4202-1232]{Qiaoya Wu}
\email{qiaoyaw2@illinois.edu}
\affiliation{Department of Astronomy, University of Illinois at Urbana-Champaign, Urbana, IL 61801, USA}

\author[0000-0003-1659-7035]{Yue Shen}
\email{shenyue@illinois.edu}
\affiliation{Department of Astronomy, University of Illinois at Urbana-Champaign, Urbana, IL 61801, USA}
\affiliation{National Center for Supercomputing Applications, University of Illinois at Urbana-Champaign, Urbana, IL 61801, USA}

\begin{abstract}
The Dark Energy Spectroscopic Instrument (DESI) 
survey will provide optical spectra for $\sim 3$ million 
quasars. Accurate redshifts for these quasars will facilitate a broad range of science applications. Here we provide improved systemic redshift estimates for the $\sim 95$k quasars included in the DESI Early Data Release (EDR), based on emission-line fits to the quasar spectra. The majority of the DESI pipeline redshifts are reliable. However, $\sim 19\%$ of the EDR quasars have pipeline redshifts that deviate from our new redshifts by $>500\,{\rm km\,s^{-1}}$. We use composite quasar spectra to demonstrate the improvement of our redshift estimates, particularly at $z>1$. These new redshifts are available at \url{https://github.com/QiaoyaWu/DESI_EDR_qsofit/blob/main/DESI_EDR_Aug29_redshift_only.fits}.   
\end{abstract}


\section{Introduction and Conclusions}\label{sec:intro}

Wide-field spectroscopic surveys such as the Dark Energy Spectroscopic Instrument \citep[DESI;][]{Levi_etal_2013_DESI} survey will facilitate the statistical studies of quasars and cosmology with unprecedented large spectroscopic samples. 
DESI began its science observations in December 2020 and is expected to obtain optical spectra for more than three million quasars during its 5-year operations. The DESI spectroscopic data from the first five-month Survey Validation observations are included in the Early Data Release \citep[EDR;][]{DESI_EDR}, which includes spectra for 1.8 million unique targets, among which $\sim 95$~k are classified as broad-line quasars.

Spectra in EDR are accessible in two primary categories: tile-based coadded spectra that combine targets across multiple exposures of the same tile (but not across different tiles even if the same target was observed on multiple tiles); and HEALPixel-based coadds that combine exposures by the HEALPix location on the sky (in both cases, data are not combined across surveys and programs). To reduce duplicated targets in our analysis, we use the HEALPix-based coadded spectra to perform spectral analysis.
We start from the HEALPixel-based redshift value-added catalog, which has additional columns and fixes compared to the original catalog. We restrict to spectra with the fiber placed on target (column OBJTYPE==TGT) and no errors or flags due to instrument problems or spectroscopic pipeline issues (column ZWARN==0). We identified 94,678 spectra classified by the pipeline as broad-line quasars (column SPECTYPE==QSO).

We measure spectral properties for these $\sim 95$k EDR quasars following the practice in earlier work \citep[e.g.,][]{Shen_etal_2011, Shen_etal_2019b, Wu&Shen2022}. Each quasar spectrum is fit with a global continuum+emission lines model, using the public code {\tt PyQSOFIT} \citep{Guo_etal_2018}, with custom adjustments to improve the robustness of spectral measurements. The detailed methodology of our spectral fitting procedure is described in our earlier work \citep{Wu&Shen2022}. 

To derive a systemic redshift from the spectral fit, we follow our earlier work \citep{Shen_etal_2016b, Wu&Shen2022}. Each line included in the fit provides an estimate of the systemic redshift, after accounting for the average intrinsic velocity shifts of different lines, and their luminosity dependence for two high-ionization lines: \CIV and \SiIV. We estimate the systematic redshift $z_{\rm sys, line}$ for each individual line (narrow \halpha, 
broad \hbeta, \OIII, \CaII, \OII, \MgII, \CIII, \CIV and \SiIV) with the quality cut on the line flux detection at $>2\sigma$ and spectral coverage of the line. We then derive a mean systemic redshift $z_{\rm sys}$ from all available $z_{\rm sys, line}$ weighted by their individual uncertainties, after rejecting outlier $z_{\rm sys, line}$ values that are $>3$ times the median absolute deviation from the mean.

For $\sim80\%$ of the DESI EDR sample, the redshift difference between $z_{\rm EDR}$ and our $z_{\rm sys}$ is small ($v_z= |\Delta V|\equiv|c\Delta z/(1+z_{\rm sys})|<500\,{\rm km\, s^{-1}}$), indicating that the DESI redshift pipeline performs well overall. However, 13.2\%, 3.2\%, and 2.2\% of the quasars have redshift differences $|\Delta V|\in[500, 1000]$, $|\Delta V|\in[1000, 1500]$, and $|\Delta V|>1500\,{\rm km\, s^{-1}}$, respectively. This fraction of redshift discrepancy generally increases with redshift, and reaches as high as $\sim 50\%$ at $z\sim 3$ (though the number of DESI quasars peaks at $z\sim 2$). 

\begin{figure*}
\centering
    \includegraphics[width=\linewidth]{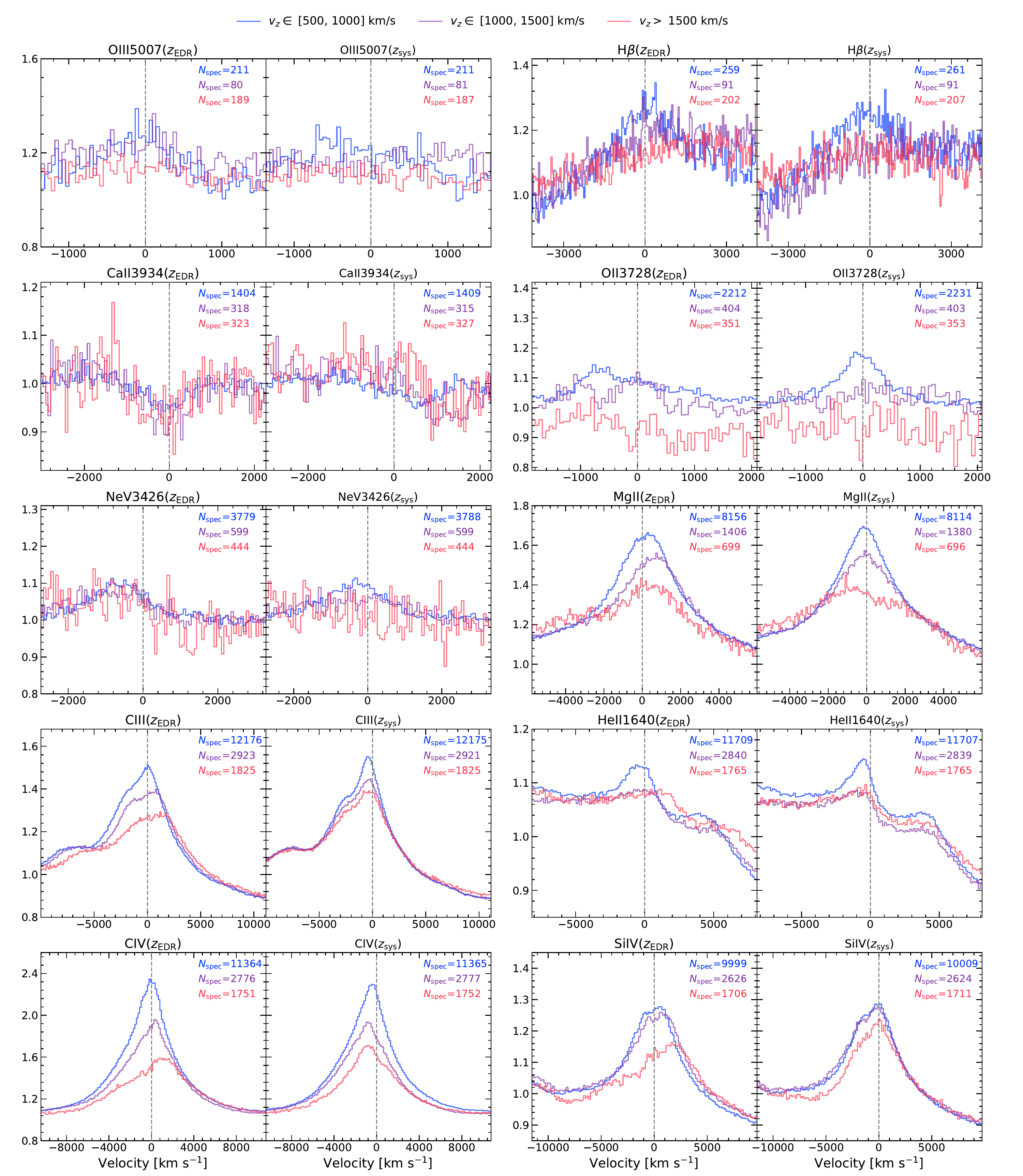}
    \caption{Median composite spectra around several emission/stellar absorption lines of DESI EDR quasars with large discrepancies between the EDR redshift and the new systemic redshift, generated using the EDR redshifts (left panels) and the improved systemic redshift (right panels). The number of quasars contributing to the coadds for each particular line is listed in the upper right corner of each. The new systemic redshifts for high-redshift quasars ($z>1$) perform much better than $z_{\rm EDR}$ in the statistical sense. 
    }
    \label{fig:comp_spec}
\end{figure*}

Figure \ref{fig:comp_spec} compares the median composite spectra for subsets of DESI EDR quasars with large differences between $z_{\rm EDR}$ and $z_{\rm sys}$. In this comparison, for high-$z$ quasars with rest-frame UV line coverage in DESI spectra, our new redshifts are significantly improved over $z_{\rm EDR}$. The broad line profiles in the composite spectrum are sharper, and their peak locations are more consistent with their expected velocity shifts from the systemic velocity \citep{Shen_etal_2016b}. For example, the \CIV\ line is expected to be blueshifted from the systemic velocity by hundreds of ${\rm km\,s^{-1}}$, while the peak of \MgII\ is expected to be close to systemic. The \CIII\ and \SiIV\,1400\AA\ complexes are also better defined using our $z_{\rm sys}$.

At low redshift where the DESI spectra cover \halpha\ or \hbeta+\OIII, their redshifts produced by the DESI pipeline are generally accurate, and only a small fraction of quasars show discrepant redshifts. The weak line strength in both composite spectra may be a result of low signal-to-noise ratio for this subset of quasars that show large redshift discrepancies (many of them indeed have noisy spectra and our spectral fits are problematic). In fact, given the low SNR of these spectra, the pipeline redshifts $z_{\rm EDR}$ based on template fitting are better than our line-based $z_{\rm zsys}$ on average. We are improving our emission-line-based redshifts for low-SNR spectra.

In addition, we find many EDR quasars have catastrophic redshift failures in the DESI pipeline (e.g., a UV broad line is mistaken as \halpha\ by the pipeline). Because we used the DESI pipeline redshift as the input redshift in our spectral fitting, our $z_{\rm sys}$ value will also be incorrect for these quasars. In future work, we will correct these catastrophic redshifts using the combination of automatic flagging and visual inspection, as did for the SDSS-DR16 catalog \citep{Wu&Shen2022}. 


To summarize, we performed spectral fits for the $\sim 95$k DESI EDR quasars and provide improved systemic redshifts based on emission lines. This initial investigation on DESI quasar pipeline redshifts does not include visual inspection and correction for catastrophic redshift failures by the DESI pipeline. In future work, we will present a more comprehensive compilation of all spectral properties from our spectral fits and improved systemic redshifts for DESI quasars.

\bibliography{refs}{}
\bibliographystyle{aasjournal}

\end{document}